\documentclass[reprint,aps,pra,twocolumn,superscriptaddress,showpacs,floatfix,longbibliography,citeautoscript]{revtex4-1}

\usepackage[dvips]{graphicx}

\usepackage{color}

\usepackage{latexsym}

\usepackage{amsmath}

\usepackage{amsthm}

\usepackage{amsfonts}

\usepackage{amssymb}

\usepackage{mathptmx}

\usepackage{subfigure}

\usepackage{mathrsfs}

\usepackage{hyperref}

\hypersetup{colorlinks=true}

\usepackage[all]{hypcap}

\usepackage{gensymb}

\begin{document}
\title{Unravelling the complex magnetic structure of multiferroic pyroxene NaFeGe$_2$O$_6$: A combined experimental and theoretical study}
\author{Lei Ding}
\email[]{lei.ding.ld@outlook.com; lei.ding@sftc.ac.uk}
\affiliation{ISIS Facility, Rutherford Appleton Laboratory, Harwell Oxford, Didcot OX11 0QX, United Kingdom}
\author{Pascal Manuel}
\affiliation{ISIS Facility, Rutherford Appleton Laboratory, Harwell Oxford, Didcot OX11 0QX, United Kingdom}
\author{Dmitry D. Khalyavin}
\affiliation{ISIS Facility, Rutherford Appleton Laboratory, Harwell Oxford, Didcot OX11 0QX, United Kingdom}
\author{Fabio Orlandi}
\affiliation{ISIS Facility, Rutherford Appleton Laboratory, Harwell Oxford, Didcot OX11 0QX, United Kingdom}
\author{Alexander~A.~Tsirlin}
\email[]{altsirlin@gmail.com}
\affiliation{Experimental Physics VI, Center for Electronic Correlations and Magnetism, Institute of Physics, University of Augsburg, 86135 Augsburg, Germany}
\date{\today}

\begin{abstract}
Magnetic order and the underlying magnetic model of the multiferroic pyroxene NaFeGe$_2$O$_6$ are systematically investigated by neutron powder diffraction, thermodynamic measurements, density-functional band-structure calculations, and Monte-Carlo simulations. Upon cooling, NaFeGe$_2$O$_6$ first reveals one-dimensional spin-spin correlations in the paramagnetic state below about 50\,K, uncovered by magnetic diffuse scattering. The sinusoidal spin-density wave with spins along the $a$-direction sets in at 13\,K, followed by the cycloidal configuration with spins lying in the $(ac)$ plane below 11.6\,K. Microscopically, the strongest magnetic coupling runs along the structural chains, $J_1\simeq 12$\,K, which is likely related
to the one-dimensional spin-spin correlations. The interchain couplings $J_2\simeq 3.8$\,K and $J_3\simeq 2.1$\,K are energetically well balanced and compete, thus giving rise to the incommensurate order in sharp contrast to other transition-metal pyroxenes, where one type of the interchain couplings prevails. The magnetic model of NaFeGe$_2$O$_6$ is further completed by the weak single-ion anisotropy along the $a$-direction. Our results resolve the earlier controversies regarding the magnetic order in NaFeGe$_2$O$_6$ and establish relevant symmetries of the magnetic structures. These results, combined with symmetry analysis, enable us to identify the possible mechanisms of the magnetoelectric coupling in this compound. We also elucidate microscopic conditions for the formation of incommensurate magnetic order in pyroxenes.
\end{abstract}


\maketitle

\section{Introduction} 
Spin-driven multiferroics, where significant coupling between magnetic order and electric polarization emerges due to simultaneous symmetry breaking induced by the incommensurate magnetic structure~\cite{Khomskii2009}, have drawn a great deal of attention in recent years. Several theoretical models, such as inverse Dzyaloshinsky-Moriya (DM) (or spin-current) model~\cite{Mostovoy2006, Katsura2005} and spin-dependent $p-d$ orbital hybridization~\cite{Arima2007}, have been put forward to explain this fascinating phenomenon. Although the inverse DM model essentially captured the behavior of many multiferroic materials, such as TbMnO$_3$~\cite{Kimura2003nature} and AgFeO$_2$~\cite{Terada2012} with the cycloidal spin configuration and propagation vector lying in the spin plane, it failed to account for multiferroicity in systems with proper-screw magnetic symmetry. More recently, the mechanism of ferroaxiality of the crystal structure was proposed to explain the experimentally observed multiferroic properties of Cu$_3$Nb$_2$O$_8$~\cite{Johnson2011}, CaMn$_7$O$_{12}$~\cite{Johnson2012}, and RbFe(MoO$_4$)$_2$~\cite{Hearmon2012}, where proper-screw magnetic structures with spin plane perpendicular to the propagation vector have been found. In addition, Kaplan and Mahanti~\cite{Kaplan2011} have shown that the extended inverse DM effect may contribute to the microscopic electric polarization in both cycloid and proper-screw helical systems. This observation was used to account for the multiferroicity in some of the delafossites~\cite{Terada2014Review}. 

As one of the main components of the Earth's crust and upper mantle, pyroxenes with the chemical formula AMT$_2$O$_6$ (A = mono- or divalent metal, M = transition metal, T = Ge or Si) have gained renewed attention of condensed-matter physicists, since recently a number of magnetic pyroxenes were found to show multiferroicity or magnetoelectric effect~\cite{Jodlauk2007, Streltsov2008, Kim2012, Ackermann2015,Ding2016_1,Ding2016_2}. Subsequent investigation showed that only NaFeGe$_2$O$_6$~\cite{Kim2012}, SrMnGe$_2$O$_6$~\cite{Ding2016_1}, and the mineral aegirine~\cite{Jodlauk2007} are truly multiferroic. 

\begin{figure*}[t]
\includegraphics[width=0.7\linewidth]{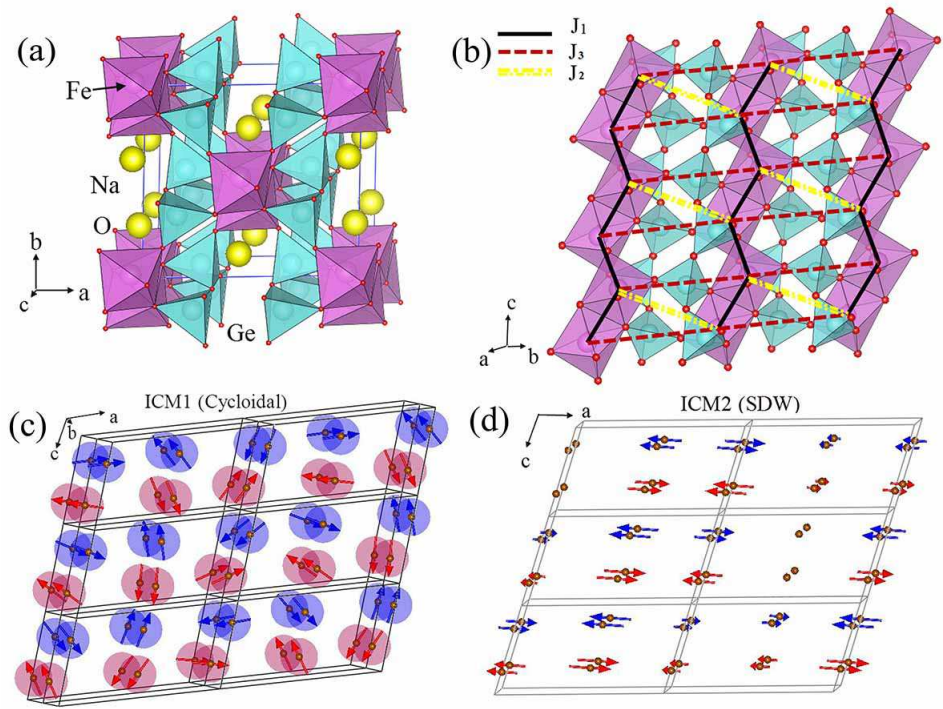}
\caption{(color online) (a) and (b) The crystal structure of NaFeGe$_2$O$_6$ with projections along different directions. The essential exchange interactions within and between the chains are highlighted. Schematic drawings of the cycloidal spin configuration (c) and spin- density wave (SDW) (d).}  \label{fig:1}
\end{figure*}

NaFeGe$_2$O$_6$ crystallizes in the space group $C2/c1'$. The zig-zag chains of edge-sharing FeO$_6$ octahedra are bridged by corner-linked GeO$_4$ tetrahedral chains (Fig.~\ref{fig:1}). This structural one-dimensionality gives rise to the broad maximum in the magnetic susceptibility around 35\,K. Two consecutive magnetic transitions at $T_{N2}=13$\,K and $T_{N1}=11.6$\,K, respectively, were identified through the specific-heat measurements~\cite{Redhammer2011,Drokina2011}. The second transition is accompanied by the formation of spontaneous electric polarization confirmed by electric polarization measurements on both powders and single crystals~\cite{Kim2012, Ackermann2015}.

Neutron diffraction studies suggest incommensurate nature of the magnetic order in NaFeGe$_2$O$_6$~\cite{Drokina2011, Redhammer2011}. However, even the periodicity of the magnetic structure remains controversial. Two different propagation vectors, $\mathbf k=(0.3357, 0, 0.0814)$~\cite{Drokina2011} versus $\mathbf k=(0.323, 1.0, 0.08)$~\cite{Redhammer2011}, were reported by different groups. These vectors can not be transformed into each other, because $(0,1,0)$ is not a reciprocal translation in the presence of $C$-centering. The magnetic structure below $T_{N1}$ was determined to be cycloidal, whereas the magnetic structure of the intermediate phase formed between $T_{N1}$ and $T_{N2}$ has not been reported to date. The controversial information on the magnetic structure, along with the absence of any established microscopic magnetic model, hinder further work on NaFeGe$_2$O$_6$ and curtail our understanding of the multiferroicity of this compound.

In the following, we revisit the magnetic structure of NaFeGe$_2$O$_6$ and establish the microscopic magnetic model. We demonstrate that the well-tuned balance between the interchain couplings gives rise to the incommensurate order and renders NaFeGe$_2$O$_6$ different from the majority of pyroxenes that feature collinear and commensurate magnetic structures. We further resolve the intermediate-temperature magnetic structure between $T_{N1}$ and $T_{N2}$ as the spin-density wave caused by the weak single-ion anisotropy of Fe$^{3+}$. We finally discuss implications of our results for the multiferroic behavior, as well as microscopic conditions for the formation of incommensurate magnetic order in pyroxenes.

\section{Methods}
Polycrystalline NaFeGe$_2$O$_6$ was synthesized by a solid-state reaction. The stoichiometric mixture of reagent-grade Na$_2$CO$_3$, Fe$_2$O$_3$, and GeO$_2$ was ground in an agate mortar and pelletized. The pellets were placed into alumina crucibles and heated in air at 1273 K for 100 hours and cooled down to room temperature. Intermediate regrinding and reheating were performed in order to improve the purity of the sample. 

The temperature-dependent magnetic susceptibility was measured using a SQUID magnetometer (Quantum Design, MPMS-7T). The dc magnetic susceptibility was recorded from 2 to 350 K in zero-field-cooled (ZFC) and field-cooled (FC) procedures in a magnetic field of 1 T. The specific-heat measurement was carried out using a relaxation technique with a Quantum Design Physical Property Measurement System (PPMS) in the temperature range of 2-300 K on cooling. The pelletized sample was mounted on a sample platform with Apiezon N-grease for better thermal contact. 

Temperature-dependent powder x-ray diffraction (XRD) data were collected with a RIGAKU Smartlab diffractometer in the high-resolution parallel beam mode using a Ge (220)$\times$2 monochromator for $Cu K \alpha_1$ radiation and Oxford Phenix cold stage giving access to sample temperatures as low as 12 K. The neutron powder diffraction (NPD) data were collected at the ISIS pulsed neutron and muon facility of the Rutherford Appleton Laboratory (UK), on the WISH diffractometer located at the second target station \cite{Chapon2011}. A powder sample ($\sim 4.1$ g) was loaded into a 6 mm diameter cylindrical vanadium can and measured in the temperature range of 1.5 - 150 K using an Oxford Instrument Cryostat. 
The data at 1.5, 20, 50, 100 and 150 K were collected for 1 hour, and typical scans between these temperatures were carried out with an exposition time of 30 minutes with steps of 1 K in the temperature range of 2-10 K and 0.2 K for measurements between 10 and 15 K. 

Rietveld refinements of the crystal and magnetic structures were performed using the Fullprof program \cite{Fullprof1993} against the data measured in the detector banks at the average $2\theta$ values of 58$^{\circ }$, 90$^{\circ }$, 122$^{\circ }$, and 154$^{\circ }$, each covering 32$^{\circ }$ of the scattering plane. Group-theoretical calculations were done using ISODISTORT \citep{ISODISTORT} and Bilbao Crystallographic Server (Magnetic Symmetry and Applications \cite{MAGNDATA}) software.

Magnetic exchange couplings were analyzed using density-functional (DFT) band-structure calculations performed in the \texttt{FPLO}~\cite{fplo} and \texttt{VASP}~\cite{vasp1,vasp2} codes. Perdew-Burke-Ernzerhof flavor of the exchange-correlation potential was chosen~\cite{pbe96}. The $k$ mesh with up to 64 points in the symmetry-irreducible part of the first Brillouin zone was used and proved sufficient for the full convergence with respect to the number of $k$-points. Correlation effects in the Fe $3d$ shell were taken into account on the mean-field level via the DFT+$U$ procedure with the on-site Coulomb repulsion $U_d=6-8$\,eV and Hund's exchange $J_d=1$\,eV~\cite{vasiliev2016,tsirlin2017}.

Exchange couplings $J_{ij}$ enter the spin Hamiltonian
\begin{equation}
 H = \sum_{\langle ij\rangle}J_{ij}\mathbf S_i\mathbf S_j+\sum_i A_i (S_i^z)^2
\label{eq:spin}\end{equation}
where $S=\frac52$ and the summation is over bonds $\langle ij\rangle$. The values of $J_{ij}$ were obtained by a mapping procedure using energies of collinear spin configurations~\cite{xiang2013}. A similar mapping procedure for orthogonal spin configurations yields magnetic anisotropy parameters $A_i$ when spin-orbit (SO) coupling is included within the DFT+$U$+SO approach. 

Thermodynamic properties of the resulting spin model were analyzed by classical Monte-Carlo simulations using the \texttt{spinmc} algorithm of the \texttt{ALPS} package~\cite{alps}. Finite $L\times L\times L$ lattices with $L\leq 8$ and periodic boundary conditions were used.

\section{Results}

\subsection{Magnetic properties}

\begin{figure}
\centering
\includegraphics[width=0.8\linewidth]{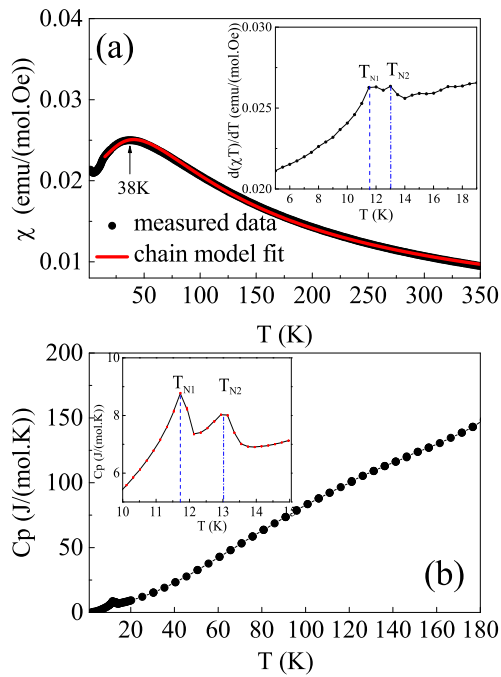}
\caption{ (a) Temperature dependence of the magnetic susceptibility of NaFeGe$_2$O$_6$ in a magnetic field H=1 T. The inset shows Fisher's heat capacity $d(\chi T)/dT$. (b) Heat capacity ($C_p$) for NaFeGe$_2$O$_6$. Two magnetic transitions at T$_{N1}$ and T$_{N2}$ are marked in the inset.}\label{fig:2}
\end{figure}

The temperature dependence of the magnetic susceptibility of NaFeGe$_2$O$_6$ measured in a magnetic field of 1\,T is shown in Fig.~\ref{fig:2} (a). An obvious broad maximum at $\sim$ 38 K resembles the behavior of a linear-chain Heisenberg antiferromagnet, in agreement with the chain-like structural features~\cite{Nenert2015}. In fact, similar low-dimensional features have also been observed in other pyroxenes, such as NaCrGe$_2$O$_6$~\cite{Nenert2009}. With further decreasing temperature, a drop around 11.6 K occurs. As marked by the dashed and dashed-dotted lines in the inset, two distinct magnetic transitions at T$_{N1}$ = 11.6 K and T$_{N2}$ = 13 K can be clearly seen in the Fisher's heat capacity $d(\chi T)/dT$, suggesting two magnetically ordered states. This result is consistent with the previous studies~\cite{Kim2012,Drokina2011}. 

Experimental magnetic susceptibility was fitted with the Curie-Weiss law between 200 and 350 K. This yields an effective moment $\mu_{eff}$ = 6.16(8)$\mu_B$, consistent with the calculated spin-only value of 5.92 $\mu_B$ for the Fe$^{3+}$ cations in the high-spin state, in agreement with the previous report~\cite{Kim2012}. The negative Weiss temperature of $\Theta=-117(1)$\,K indicates predominant antiferromagnetic interactions and reveals a considerable reduction in the N\'eel temperature, $\Theta/T_N\simeq 10$, which may be due to the low-dimensionality and/or frustration. 

In order to further characterize these magnetic phase transitions, we performed measured heat capacity of NaFeGe$_2$O$_6$ showin in Figure \ref{fig:2} (b). The two successive cusps at 11.6 and 13 K are indicative of two magnetic phase transitions, in good agreement with our magnetic susceptibility data. No apparent anomaly can be observed around 38 K, implying that the broad maximum at ~38 K should be attributed to short-range magnetic correlations. 

\begin{figure}[t]
\centering
\includegraphics[width=0.8\linewidth]{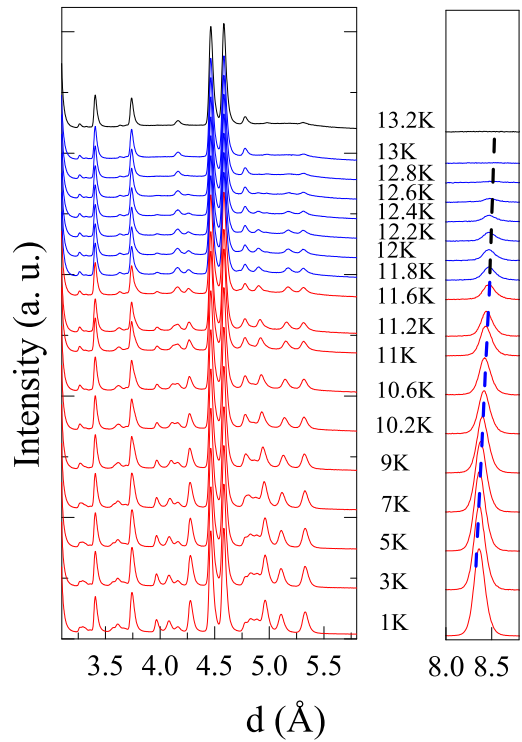}
\caption{ Temperature dependence of the neutron powder diffraction data of NaFeGe$_2$O$_6$. Black, blue, and red reflections correspond to the paramagnetic, ICM2, and ICM1 phases, respectively}\label{fig:3}
\end{figure}

\subsection{Neutron diffraction}

\begin{figure}
\centering
\includegraphics[width=0.8\linewidth]{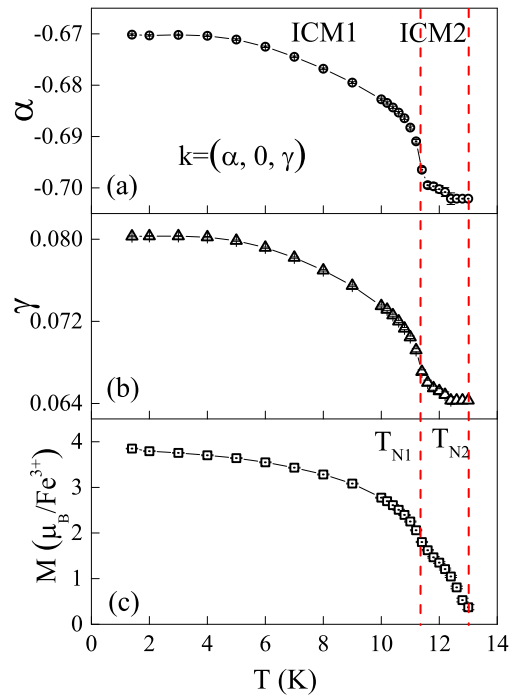}
\caption{Temperature dependence of the $\alpha$ and $\gamma$ components of the magnetic propagation vector ((a) and (b)) and the refined magnetic moment (c).}\label{fig:4}
\end{figure}

According to our temperature-dependent x-ray diffraction and the WISH backscattering data collected in the temperature range of 1.5-150 K, NaFeGe$_2$O$_6$ crystallizes with the $C2/c1'$ symmetry and has no symmetry change down to 1.5 K. At 150 K, the lattice parameters are: $a=10.0092(1)$\,\r A, $b=8.9124(1)$\,\r A, $c=5.50895(5)$\,\r A, $\beta=107.5189(9)^{\circ}$. Magnetic Bragg reflections appear below T$_{N2}$ = 13 K (ICM2 phase) in the NPD data, as shown in Fig.~\ref{fig:3}, and they can be indexed by an incommensurate propagation vector \textbf{k} $\backsimeq$ ($\alpha$, 0, $\gamma$) with $\alpha=-0.6999(8)$ and $\gamma=0.0649(2)$ at 12.2 K. The value of \textbf{k} shows a slightly temperature-dependent behavior, as indicated in Fig.~\ref{fig:4}. On further cooling, additional magnetic reflections appear below T$_{N1}$ = 11.6 K (ICM1 phase), and the magnetic reflections exhibit an obvious temperature-dependent behavior. These reflections can also be indexed by the same incommensurate vector \textbf{k} albeit with a slightly different $\alpha$ and $\gamma$ values (Fig. \ref{fig:4} (a) and (b)). At 1.5 K, the refined \textbf{k} is $(-0.6702(1),0,0.08028(5))$. It is clear that the ICM2 phase only appears within the very narrow temperature range $11.6-13$\,K. 

Previous neutron diffraction experiments on both powder and single crystals failed to resolve this phase~\citep{Redhammer2011} The presence of magnetic Bragg reflections in our neutron diffraction data is consistent with the magnetic susceptibility and heat capacity measurements, showing the existence of two ordered magnetic states. The propagation vector of the ICM1 phase we obtained is, in fact, equivalent to the vector \textbf{k}$'$=(0.323, 1.0, 0.08) reported in Ref.~\onlinecite{Redhammer2011}. By applying a reciprocal translation ($-1$,$-1$,0), one can transform \textbf{k}$'$ into $\mathbf k=(-0.67,0,0.08)$. 

\begin{figure}
\includegraphics[width=1\linewidth]{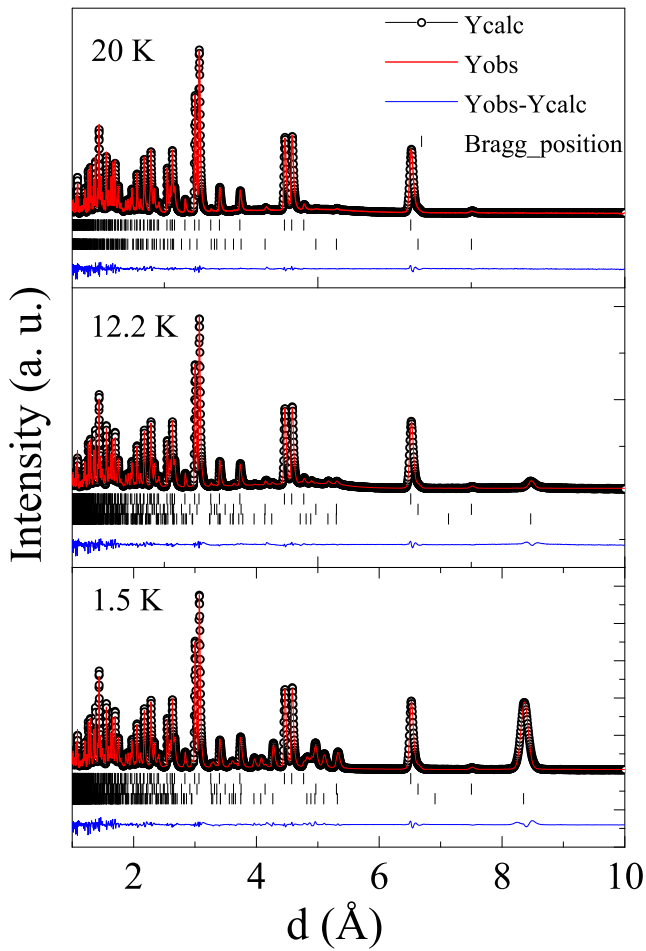}
\caption{(color online) Neutron powder diffraction patterns for NaFeGe$_2$O$_6$ at 20, 12.2 and 1.5 K. The nuclear reflections are denoted by upper tick marks. The reflections marked in the second line belong to a impurity phase Na$_{4}$Ge$_{9}$O$_{20}$  with the weight fraction of 1.52(2)\%. The lowest tick marks show magnetic phase (in case of 12.2 K and 1.5 K). } \label{fig:5}
\end{figure}    

Symmetry analysis was performed in order to determine the magnetic structures of NaFeGe$_2$O$_6$. Starting with the parent space-group $C2/c 1'$ and propagation vector \textbf{k} $\backsimeq$ ($\alpha$, 0, $\gamma$) in the B plane of the Brillouin zone, two active magnetic irreducible representations, mB1 and mB2, as well as their corresponding subgroups were obtained using ISODISTORT. For the ICM1 phase, we found that the magnetic superspace group $Cc1'(\alpha,0,\gamma)0s$ (basis={($-$1,0,0,0),(0,$-$1,0,0),(0,0,$-$1,0),(0,0,0,1)}, origin=(0,0,0,0)), generated from the single active mB1 irreducible representation, can be adopted to describe the magnetic structure. Such a symmetry fixes the phase difference between atoms Fe1 $(0,y,0.25)$ and Fe2 $(0,-y,0.75)$ at (1+$\gamma$)*$\pi$. The magnetic structure refinement at 1.5 K was carried out by taking into account this symmetry constraint. The final refinement is shown in Fig. \ref{fig:5} arriving at the cycloidal configuration with  magnetic moments in the $(ac)$ plane. The refined total magnetic moment at 1.5 K is 3.857(8) $\mu_B$, considerably smaller than 5\,$\mu_B$ expected for $S=\frac52$ of Fe$^{3+}$. In fact, this value is very close to the total magnetic moment with 4.09(4) $\mu_B$ refined from the single-crystal experiment of Ref.~\cite{Redhammer2011}. Such a reduction, observed very often in cycloidal spin systems, is likely a consequence of spin fluctuations and covalency. The magnetic symmetry, as represented in Fig. \ref{fig:1}, breaks the inversion symmetry and preserves the mirror-plane symmetry perpendicular to the unique $b$ axis, leading to the magnetic point group $m 1'$ which allows the existence of a ferroelectric polarization. Indeed this magnetic symmetry corroborates the earlier observations of multiferroicity~\cite{Kim2012, Ackermann2015}. 
   
\begin{figure}
\includegraphics[width=0.85\linewidth]{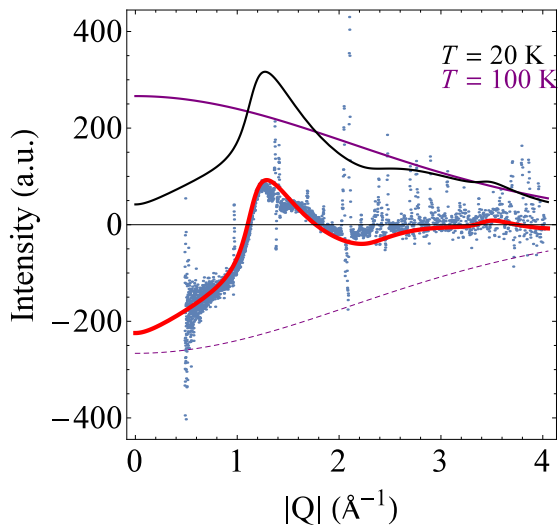}
\caption{(color online) The difference between neutron diffraction patterns at 20 and 100 K and the fitted curve based on the 1D model (equation \ref{eq: 1}).} \label{fig:6}
\end{figure}
    
The magnetic symmetry for the ICM2 phase belongs to the same irreducible representation, but with a distinct magnetic order parameter direction $(a,0)$. This corresponds to the magnetic superspace group $C2/c1'(\alpha,0,\gamma)00s$ (basis={($-$1,0,0,0),(0,$-$1,0,0),(0,0,$-$1,0),(0,0,0,1)}, origin=(0,0,0,0)), which conserves the inversion symmetry and the twofold screw axis. We found that a sinusoidally modulated magnetic structure is suitable to refine our neutron data at 12.2 K. The refinement leads to the spin moment of 1.55(2) $\mu_B$ along the $a$ axis. The final refined neutron diffraction pattern is shown in Fig. \ref{fig:5} and the corresponding magnetic configuration is illustrated in Fig. \ref{fig:1}. One can immediately see that it does not break the space inversion and gives rise to a centrosymmetric magnetic point group $2/m 1'$. Such a magnetic structure cannot lead to any long-range electric polarization, in agreement with the previous polarization measurements. The temperature-dependent ordered moment of NaFeGe$_2$O$_6$ is shown in Fig. \ref{fig:4} (c), where the magnetic moment for the ICM2 phase (SDW) is taken as a quadratic mean of the refined moment.  
   
\begin{figure}
\includegraphics[width=0.8\linewidth]{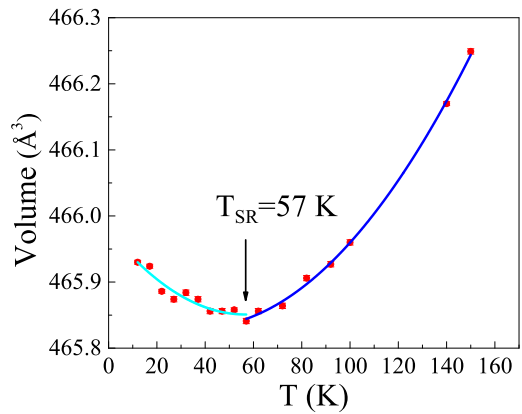}
\caption{(color online) Temperature dependence of the lattice volume of NaFeGe$_2$O$_6$ from variable-temperature XRD. } \label{fig:7}
\end{figure} 

Having resolved the long-range magnetic order in NaFeGe$_2$O$_6$, we now look into the short-range order above $T_{N2}$. As shown in Fig.~\ref{fig:6}, the magnetic diffuse scattering in NaFeGe$_2$O$_6$ extracted from the difference of the neutron diffraction data collected at 20 K and 100 K shows a maximum around $d=5.1$ \AA (~1.4 \AA$^{-1}$), signaling the presence of short-range magnetic correlations. The feature that sharply rises at low $Q$ and gradually decreases toward high $Q$ is characteristic of one-dimensional spin-spin correlations expected within the structural chains of NaFeGe$_2$O$_6$. In the family of magnetic pyroxenes, the presence of one-dimensional correlations has been evidenced in CaMnGe$_2$O$_6$ through the analysis of neutron diffuse scattering data based on an analytical one-dimensional antiferromagnetic (AFM) model \cite{Ding2016_2}, 
\begin{align}
S(Q)=f(Q)^2\,{\displaystyle \sum_{i}\langle S_{0}S_{i}\rangle\frac{\sin(QR_{i})}{QR_{i}}} \label{eq: 1}
\end{align} 
where $f(Q)$ is the magnetic form factor of Fe$^{3+}$ in the dipole approximation, and R$_i$ represents the distance between the sites along the chain. The exponential decrease and the AFM spin-spin correlations $<S_{0}S_{i}>$ with the distance $d_{i}$ and correlation length $\xi$ are expressed as
\begin{align}
\langle S_{0}S_{i}\rangle=(-1)^{i}S^{2}\exp\left(-\frac{d_{i}}{\xi}\right)
\end{align}

We fitted such a model against the experimental data, with the best fit shown in Fig \ref{fig:6}. The correlation length of 8.0 $\pm$0.4 \AA  stands for the short-range magnetic correlations along the $c$ axis. In fact, the onset temperature of the 1D spin-spin correlations is likely higher than 38 K (the position of the magnetic susceptibility maximum), as weak diffuse scattering is still present at 50 K. Another evidence for the 1D spin correlations above $T_{N2}$ is obtained from thermal expansion. As shown in Fig. \ref{fig:7}, the temperature-dependent lattice volume of NaFeGe$_2$O$_6$ refined from the XRD data exhibits apparent negative thermal expansion below 57\,K. This anomaly can be attributed to the magnetostriction effect related to the short-range magnetic order in 1D~\cite{Ding2016_2}.  

\subsection{Mean-field analysis}

\begin{table}
\caption{Contributions to the exchange matrix from the spin exchange paths between the atom sites Fe1 (0, 0.9036(2), 0.25) and Fe2 (0, $-$0.0964(2), 0.75) in a primitive setting. \label{tab:1}}
\begin{tabular}{cccccccccc}
\hline
\hline
\multicolumn{1}{c}{S$_i$}    & &  \multicolumn{1}{c}{S$_j$}   & & \multicolumn{1}{c}{d(Fe-Fe)}  & & \multicolumn{1}{c}{ \textbf{R}} & & Contribution to $\xi_{ij}$ \\
 \hline 
 Fe1  & & Fe1 &  & 6.69 \AA &  &  (1,0,0)   & & $J_3e^{-ik_x}$    \\  
   & &  &  &   &  &  (-1,0,0)    & & $J_3e^{ik_x}$   \\  
    & &  &  &  &  &  (0,1,0)    & &   $J_3e^{-ik_y}$  \\  
      & &  &  &  &  &  (0,-1,0) & & $J_3e^{ik_y}$      \\  
 Fe1 & & Fe2  & &  3.257 \AA  &  & (0,0,0)(0,0,1) & & $J_1(1+e^{-ik_z})$           \\
  & &   & &   5.63 \AA  &  & (-1,0,0)(0,-1,1) & & $J_2(e^{ik_x}+e^{i(k_y-k_z)})$           \\
   Fe2 & & Fe1 &  & 3.257 \AA  & &   (0,0,0)(0,0,-1) & & $J_1(1+e^{ik_z})$ \\
      & &  &  & 5.63 \AA   & &   (1,0,0)(0,1,-1)    & & $J_2(e^{-ik_x}+e^{i(-k_y+k_z)})$    \\
  Fe2 & & Fe2 &  & 6.69 \AA &  &  (1,0,0) & &  $J_3e^{-ik_x}$   \\
   & &  &  &  &  &  (-1,0,0) & & $J_3e^{ik_x}$ \\
   & &  &  &  &  &  (0,1,0)   & & $J_3e^{-ik_y}$  \\
   & &  &  &  &  &  (0,-1,0)  & & $J_3e^{ik_y}$  \\
\hline
\hline
\end{tabular}
\end{table}

As we confirm by the direct \textit{ab initio} analysis in Sec.~\ref{sec:dft}, the crystal structure of NaFeGe$_2$O$_6$ hosts three exchange couplings. $J_1$ runs along the chains of the FeO$_6$ octahedra, whereas $J_2$ and $J_3$ couple these chains into the 3D network. In the following, we 
use the mean-field and classical spin approximation that proved efficient in previous studies~\cite{Freiser1961, Dai2005, Khalyavin2010cal}, and investigate ordered spin configurations arising from the interplay of $J_1$, $J_2$, and $J_3$. 

Consider the primitive cell and the spin Hamiltonian given by
\begin{equation}
H=\sum\limits_{i,j}\sum\limits_{R,R'} J^{RR'}_{ij}S^R_i S^{R'}_j
\end{equation} 
where $J^{RR'}_{ij}$ is the exchange interaction between the spins $S_i$ and $S_j$. We employ the method of Freiser \cite{Freiser1961} to determine the ground state. Suppose $\sigma^R_i$ represents the mean spin at site $i$ in a cell with the lattice vector $R$. Then the ordered spin configuration can be expressed in terms of the Bloch spin functions
\begin{equation}
\sigma^R_i=\sum\limits_{k}\sigma^k_ie^{-ikR},
\end{equation}
and the spin-spin interaction energy $\xi_{ij}$ between the two sites becomes
\begin{equation}
\xi_{ij}=\sum\limits_{R}J^{R}_{ij}e^{-ikR}.
\end{equation} 
The diagonalization of the quadratic part of the mean-field energy results in the eigenvalue problem
\begin{equation}
\sum\limits_j\lbrace\sum\limits_{R'}J^{R'}_{ij}e^{-ikR'}\rbrace\sigma_j=\lambda(k)\sigma_i.
\end{equation} 
The eigenvalues are inversely proportional to the possible transition temperatures, whereas the corresponding eigenvectors yield periodicities of the spin configurations. For a given set of exchange parameters, one finds the vector \textbf{k} that delivers the lowest eigenvalue of the interaction matrix. This eigenvector will usually indicate the periodicity of the first (lowest-temperature) ordered state~\cite{Khalyavin2010cal, Luttinger1946}. 

\begin{figure}
\includegraphics[width=1\linewidth]{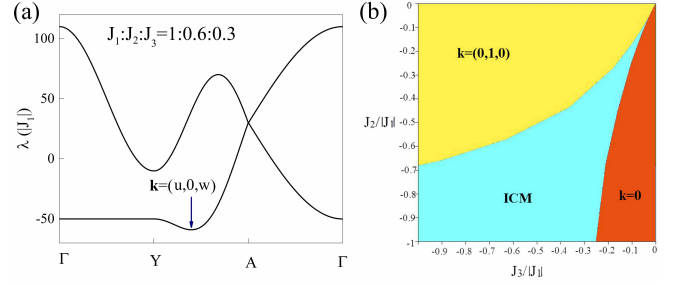}
\caption{(color online)(a) Dispersion relations for the eigenvalues of the exchange matrix with $J_1$:$J_2$:$J_3$=1:0.6:0.3 along some lines of symmetry in the first Brillouin zone of the $C2/c1'$ space group. (b) Magnetic phase diagram representing the stability of different magnetic ground states at various exchange parameters. Yellow region stands for magnetic phase characterized by \textbf k=(0,1,0) while region in red corresponds to \textbf k=0 phase. Phase colored with cyan denotes an ICM phase.} \label{fig:8}
\end{figure}

The spin-spin exchange energies associated with the matrix $\xi_{ij}$ from various spin exchange paths $J_1$, $J_2$ and $J_3$ in NaFeGe$_2$O$_6$ are summarized in Table \ref{tab:1}. We employed the program ENERMAG \cite{Khayati2001} to diagonalize the exchange matrix, and considered only the AFM case, because the couplings $J_2$ and $J_3$ are long-range and unlikely to be ferromagnetic, whereas $J_1$ is known to be AFM too~\cite{Streltsov2008,Janson2014}. The magnetic ground state depends on relative values of the exchange parameters, so we set $J_1=1$ and analyze the magnetic structure as function of $J_2/J_1$ and $J_3/J_1$ (Fig.~\ref{fig:8}). Each of the interchain couplings taken alone yields commensurate order, but of different type, $\mathbf k=0$ in the case of $J_2$ and $\mathbf k=(0,1,0)$ in the case of $J_3$. The incommensurate phase appears when both $J_2$ and $J_3$ are sizable as a result of the competition between the interchain couplings.  

The $k=0$ phase is common to transition-metal pyroxenes and has been reported, e.g., for NaCrGe$_2$O$_6$, NaCrSi$_2$O$_6$ \cite{Nenert2009, Nenert2010_2} and CaMnGe$_2$O$_6$ \cite{Ding2016_2}. It corresponds to the ferromagnetic ordering of antiferromagnetic spin chains. The $\mathbf k=(0,1,0)$ state was reported for CaM(Si,Ge)$_2$O$_6$ (M = Fe, Co, Ni), where spins are ferromagnetically coupled within chains and antiferromagnetically aligned between the chains. As for NaFeGe$_2$O$_6$, its incommensurate order is naturally ascribed to the competition between $J_2$ and $J_3$. Using the $J_1$:$J_2$:$J_3$=1:0.6:0.3 regime, we find $\mathbf k=(-0.6, 0, 0.19)$ in reasonable agreement with the experimental propagation vector from NPD. Note, however, that at this point we only analyze the periodicity of the magnetic structure and can not distinguish between, e.g., cycloid and spin-density wave. 

\subsection{Microscopic analysis}
\label{sec:dft}
For a more quantitative and material-specific description of the magnetic ordering, we proceed to the \textit{ab initio} evaluation of the exchange couplings. Several sets of crystallographic data were reported for NaFeGe$_2$O$_6$~\cite{Redhammer2011}. We performed DFT calculations for all of them and found only minor differences in the exchange parameters. The effect of the Hubbard $U_d$ is more pronounced, but it pertains to absolute values of $J$'s and does not change their hierarchy (Table~\ref{tab:exchange}).

\begin{table}
\caption{\label{tab:exchange}
Isotropic exchange couplings $J_i$ (in\,K) in NaFeGe$_2$O$_6$ as obtained from DFT+$U$ calculations with different values of the on-site Coulomb repulsion parameter $U_d$. The last line is the Curie-Weiss temperature $\Theta$ (in\,K).
}
\begin{ruledtabular}
\begin{tabular}{ccccc}
      & $d_{\rm Fe-Fe}$ & $U_d=6$\,eV & $U_d=7$\,eV & $U_d=8$\,eV \\
$J_1$ & 3.25 & 15.0 & 12.3 & 10.2 \\
$J_2$ & 5.64 & 4.5 & 3.8 & 3.1 \\
$J_3$ & 6.70 & 2.5 & 2.1 & 1.9 \\
$\Theta$ & & $-144$ & $-119$ & $-100$ \\
\end{tabular}
\end{ruledtabular}
\end{table}

By evaluating the exchange couplings in the crystallographic unit cell of NaFeGe$_2$O$_6$ (4 magnetic atoms) and in the doubled cell (8 magnetic atoms), we established that the three exchanges, $J_1-J_3$ considered above, are sufficient for the minimum microscopic descriptions, as further long-distance interactions are well below 0.1\,K. The resulting couplings are summarized in Table~\ref{tab:exchange} and can be juxtaposed with the experiment by calculating the Curie-Weiss temperature,
\begin{equation}
 \Theta=-\frac{S(S+1)}{3}\sum_i\, z_iJ_i=-\frac{35}{6}(J_1+J_2+2J_3),
\end{equation}
where $z_i$ stands for the number of couplings per Fe site. The $\Theta$ values in Table~\ref{tab:exchange} show the best agreement with the experiment for $U_d=7$\,eV that yields $J_2/J_1=0.31$ and $J_3/J_1=0.17$. On the structural level, this hierarchy follows the increase in the Fe--Fe distances. We also note that $J_2$ involves the double GeO$_4$ bridge (two tetrahedra linking the FeO$_6$ octahedra), whereas in the case of $J_3$ only a single bridge is involved. For comparison, in Cr-based pyroxenes the interactions via the double tetrahedral bridges are predominant as well~\cite{Janson2014}. 

\begin{figure}
\includegraphics[width=0.8\linewidth]{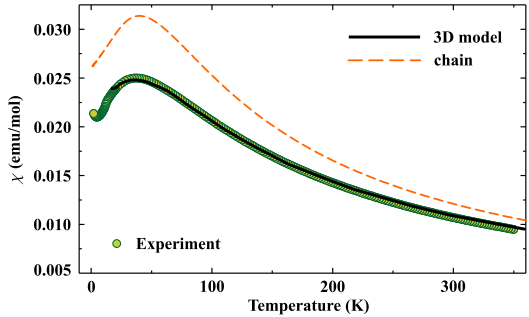}
\caption{
Fit of the magnetic susceptibility with the 3D spin model including $J_1$, $J_2$, and $J_3$, as explained in the text. The susceptibility of a spin chain with $g=2.0$ and the same $J_1$ as in the 3D fit is shown with the dashed line for reference.
} \label{fig:fit}
\end{figure}

Given the two couplings $J_1$, two couplings $J_2$, and four couplings $J_3$ per Fe site, NaFeGe$_2$O$_6$ should be far from magnetic one-dimensionality, because $J_{\rm inter}/J_{\rm intra}=(J_2+2J_3)/J_1=0.66$. On the other hand, $J_2$ and $J_3$ form triangular loops and compete (Fig.~\ref{fig:1}). This competition can also also been seen from the fact that $J_2$ and $J_3$ stabilize different types of the interchain order. The fact that $2J_3=4.2$\,K is similar to $J_2=3.8$\,K renders NaFeGe$_2$O$_6$ strongly frustrated. This frustration not only triggers the incommensurate ordering, but also introduces short-range order in the paramagnetic state, as we present below.

Regarding the long-range ordered state, simple classical minimization leads to an incommensurate state with the propagation vector $\mathbf k=(-0.675,0,0.09)$ in good agreement with the experimental $\mathbf k=(-0.67,0,0.08)$ at 1.5\,K. Monte-Carlo simulations produce magnetic susceptibility with a broad maximum and the magnetic transition taking place well below this maximum, a signature of short-range correlations above $T_N$. By keeping the $J_2/J_1$ and $J_3/J_1$ ratios from DFT and adjusting $J_1$ as well as other parameters, we arrive at the best fit with $g=1.99$ and $J_1=9.6$\,K that corresponds to the susceptibility maximum at 38\,K and $T_N\simeq 12$\,K. Note that this model features only one magnetic transition, because no anisotropy terms are involved. 

We also calculated magnetic susceptibility for a single spin chain with the same value of $J_1=9.6$\,K and $g=2.0$. As shown in Fig. \ref{fig:fit}, it reproduces the overall shape of the experimental susceptibility data but the absolute values do not match. This confirms that the susceptibility maximum in NaFeGe$_2$O$_6$ is related to the magnetic one-dimensionality, yet the interchain couplings are clearly non-negligible. 

Finally, we estimate the single-ion magnetic anisotropy. To this end, we fix spins along a given direction and rotate the reference spin in the plane perpendicular to this direction~\cite{xiang2013}. This yields $a$ as the magnetic easy axis. Placing the reference spin along $b$ and $c$ increases the energy by 0.50\,K and 0.62\,K, respectively, leading to an effective single-ion anisotropy of $A\simeq 0.09$\,K and $z=a$ in Eq.~\eqref{eq:spin}. This weak anisotropy is of similar size as in other Fe$^{3+}$ oxide compounds~\cite{tsirlin2017}. The easy-axis anisotropy naturally explains the formation of the SDW state with spins along $a$ in the ICM2 phase, because in the presence of anisotropy a collinear structure is preferred at elevated temperatures over a non-collinear one~\cite{Melchy2009}. The cycloid in the ICM1 phase features a component along the $a$ direction too, which is consistent with the calculated single-ion anisotropy.

\section{Discussion}
\begin{figure}
\includegraphics[width=0.8\linewidth]{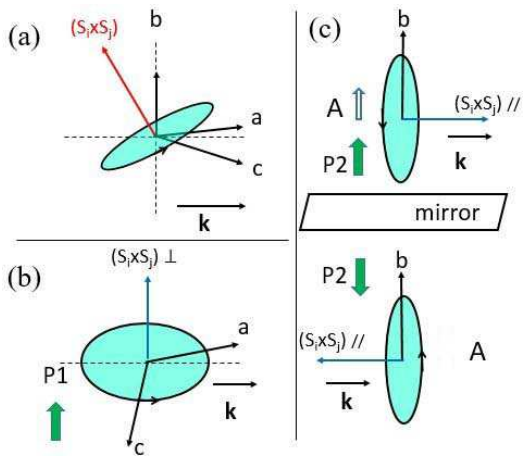}
\caption{(color online) (a) The relationship between spin helicity vector and the propagation vector $\mathbf k$ assuming the presence of the $b$-spin component. (b) The projection of spin helicity vector onto the direction perpendicular to the vector $\mathbf k$. (c) The projection of spin helicity vector onto the direction parallel to the vector $\mathbf k$ and its mirror plane related one.} \label{fig:10}
\end{figure}

Transition-metal pyroxenes show variable magnetic dimensionality and different types of the long-range order. The dimensionality changes between quasi-1D and 3D depending on the tetrahedral group~\cite{Janson2014}, whereas several flavors of commensurate and collinear long-range order were reported in previous studies~\cite{Nenert2009, Nenert2010_2,Ding2016_2}. Some of the pyroxenes show signatures of the frustration, such as the enhanced ratio $\Theta/T_N$ between the Curie-Weiss and N\'eel temperatures, but this reduction in $T_N$ is typically related to the magnetic one-dimensionality~\cite{Janson2014}. NaFeGe$_2$O$_6$ reveals a distinct microscopic scenario, where frustration is present and plays central role. The competing interactions $J_2$ and $J_3$ are well balanced and trigger incommensurate magnetic order, which is uncommon to pyroxenes. Despite the sizable interchain interactions, NaFeGe$_2$O$_6$ shows signatures of 1D magnetism above $T_{N2}$, because the chains are effectively decoupled. We note in passing that a similar microscopic scenario may be relevant to SrMnGe$_2$O$_6$, where an incommensurate magnetic structure was revealed by neutron diffraction~\cite{Ding2016_1}.

Another distinct feature of NaFeGe$_2$O$_6$ are its two consecutive magnetic transitions at $T_{N1}$ and $T_{N2}$. The majority of pyroxenes show only one magnetic transition, as expected in non-frustrated antiferromagnets. The frustration itself, the competition between $J_2$ and $J_3$, does not split the transition into two, and the presence of weak single-ion anisotropy seems to be crucial here. Similar combinations of the cycloid and SDW phases were observed in systems like Ca$_3$Co$_2$O$_6$~\cite{agrestini2008,agrestini2011} and Li$_2$NiW$_2$O$_8$~\cite{ranjith2016}, where magnetic ions bear strong single-ion anisotropy. Although Fe$^{3+}$ with its half-filled $d$-shell is by far less anisotropic than Ni$^{2+}$ or Co$^{3+}$, the anisotropy energy of less than 1\,K (and less than 1\% of the leading exchange coupling $J_1$) is already sufficient for driving similar physics. The main difference is the incommensurate and non-collinear ground-state magnetic configuration stabilized by the isotropic exchange couplings $J_i$ in NaFeGe$_2$O$_6$, whereas in systems with stronger anisotropy, commensurate and collinear states favored by the anisotropy occur.

Altogether, we resolved the earlier controversies regarding the magnetic structures of NaFeGe$_2$O$_6$, established the microscopic magnetic model of this compound, and outlined the microscopic condition for the formation of incommensurate spin states in transition-metal pyroxenes (Fig.~\ref{fig:8}). Let us now discuss the multiferroic behavior of NaFeGe$_2$O$_6$ from the symmetry perspective of the magnetic structures determined in this work. 

The magnetic superspace group $Cc1'(\alpha,0,\gamma)0s$ of the ICM1 phase breaks both spatial inversion and time reversal. This cycloidal magnetic symmetry allows electric polarization within the $(ac)$ plane (Fig. \ref{fig:10} (b)), in good agreement with the experimental observation. The polarization can be explained by the theory of the inverse DM effect or spin-current model. However, this mechanism does not account for the observation of a small polarization (less than 2\,$\mu$C/m$^2$) along the $b$-axis in a synthetic single crystal \cite{Ackermann2015}. In principle, the symmetry analysis of NaFeGe$_2$O$_6$ allows the presence of a magnetic moment along the $b$-axis and indicates that both cycloidal and proper-screw components might be present, as illustrated in Fig.~\ref{fig:10} (a). We also examined other recently developed mechanisms for explaining multiferroicity in materials showing proper-screw magnetic symmetry. The cycloidal spin arrangement (Fig. \ref{fig:10} (b)) based on $\mathbf P \propto (A \cdot \mathbf r_{ij})(\mathbf S_i \times \mathbf S_j)$ gives no electric polarization along the $b$-axis, because the mirror plane contains $\mathbf r_{ij}$ \cite{Kaplan2011, Terada2012, Terada2014, Terada2015}. In light of the ferroaxial mechanism, the proper-screw component can lead to $\mathbf P_2 \propto A \cdot \lbrace \mathbf r_{ij} \cdot (\mathbf S_i \times \mathbf S_j) \rbrace$ along the $b$-axis. However, the mirror plane perpendicular to the $b$-axis leads to the opposite spin chirality, as explained schematically in Fig. \ref{fig:10} (c), this leading to the cancellation of the electric polarization. As a result, the magnetic superspace group requires that the electric polarization can only be present in the $(ac)$ plane. A signal in the pyrocurrent measurement along the $b$-direction can be then due to a misalignment of the crystal or an impurity phase, such as hematite ($\alpha$-Fe$_2$O$_3$) and maghemite ($\gamma$-Fe$_2$O$_3$) that were identified in the crystal on which the pyrocurrent measurement of Ref.~\cite{Ackermann2015} was performed.

\section{Conclusion}

In conclusion, we present the revisited magnetic structures and associated microscopic magnetic model for NaFeGe$_2$O$_6$. This compound shows cycloid magnetic configuration below 11.6\,K preceded by a spin-density-wave state below 13\,K and a region of one-dimensional spin-spin correlations extending up to at least 50\,K. Competing interchain couplings $J_2$ and $J_3$ decouple the spin chains above $T_N$ and render magnetic order incommensurate below $T_N$. The cycloid phase is a direct result of this competition, whereas the SDW phase should form upon the presence of weak single-ion anisotropy that tends to align the spins along the $a$ direction. We report the general magnetic phase diagram of transition-metal pyroxenes, which captures all the documented magnetic
structures reported in pyroxenes so far, and argue that the electric polarization of NaFeGe$_2$O$_6$ should be confined to the $ac$-plane within the cycloid phase, whereas no electric polarization should occur within the SDW phase, which is centrosymmetric.
 
\section{Acknowledgments}

LD thanks support from the Rutherford International Fellowship Programme (RIFP). This project has received funding from the European Union's Horizon 2020 research and innovation programme under the Marie Sk\l{}odowska-Curie grant agreements No.665593 awarded to the Science and Technology Facilities Council. P. M. and D. D. K. acknowledge a support from the project TUMOCS. This project has received funding from the European Union's Horizon 2020 research and innovation programme under the Marie Sk\l{}odowska-Curie grant agreements No.645660. AAT was funded by the Federal Ministry for Education and Research through the Sofja Kovalevskaya Award of Alexander von Humboldt Foundation. We would like to thank G. Stenning and D. Nye​ for their help during our thermodynamic measurements in the Materials Characterisation Laboratory at the ISIS Neutron and Muon Source.

\end{document}